# Simple analytical expression for the peak-frequency shifts of plasmonic resonances for sensing

Jianji Yang[1], Harald Giessen[2] and Philippe Lalanne*[1]

[1] Laboratoire Photonique Numérique et Nanosciences, Institut d'Optique d'Aquitaine, Université Bordeaux, CNRS, 33405 Talence, France.
[2] 4th Physics Institute and Research Center SCoPE, University of Stuttgart, Pfaffenwaldring 57, 70550 Stuttgart, Germany.

**Abstract**
We derive a closed-form expression that accurately predicts the peak frequency-shift and broadening induced by tiny perturbations of plasmonic nanoresonators without critically relying on repeated electrodynamic simulations of the spectral response of nanoresonator for various locations, sizes or shapes of the perturbing objects. The force of the present approach, in comparison with other approaches of the same kind, is that the derivation is supported by a mathematical formalism based on a rigorous normalization of the resonance modes of nanoresonators consisting of lossy and dispersive materials. Accordingly, accurate predictions are obtained for a large range of nanoparticle shapes and sizes, used in various plasmonic nanosensors, even beyond the quasistatic limit. The expression gives quantitative insight, and combined with an open-source code, provides accurate and fast predictions that are ideally suited for preliminary designs or for interpretation of experimental data. It is also valid for photonic resonators with large mode volumes.

**Introduction.** In recent years, metallic nanoparticles have gained a lot of attention and also witnessed successful applications in various fields of nanosciences. As their near-fields support strong and highly-confined resonances, metallic nanoparticles can effectively convert local changes of refractive index into frequency shifts of the resonance.[1,2] This property has driven considerable development in sensing technologies based on localized surface plasmon resonance (LSPR) of metallic nanoparticles,[2,3] benefiting from significant advances in the detection of single metallic nanoparticles.[4,5] Thanks to their small mode volume, LSPRs are suitable for achieving detection of ultra-small refractive-index changes.[2,6-8] Even single-molecule sensitivity has been recently achieved.[9,10]

In this work, we derive an analytical formula that predicts the change $\Delta\tilde{\omega}$ of the complex-valued eigenfrequency $\tilde{\omega}$ of the LSPR, induced by a local near-field perturbation of resonant metallic nanoparticles. The derivation is motivated by the fact that the frequency shift $Re(\Delta\tilde{\omega})$ and the resonance broadening $-2Im(\Delta\tilde{\omega})$ are two quantities of fundamental importance for sensing applications.[2,3,7] Yet another motivation is that the theoretical prediction of $\Delta\tilde{\omega}$ usually relies on tedious and repeated fully-vectorial electromagnetic calculations for various parameters, such as the location, size, shape, or refractive index of the perturbation, and that a simple and intuitive approach that is accurate for *arbitrary nanoparticle sizes and shapes* may help early designs or interpretation of experimental results and is essentially absent in the literature. Finally another motivation is that the derivation of an analytical formula accurate for photonic or plasmonic nanoresonators of *arbitrary size and shape*, potentially composed of lossy and dispersive materials, is of fundamental interest and has not been already established. All these issues are addressed thanks to the introduction of the true resonance modes of lossy and dispersive resonant nanostructures and to the use of a rigorous normalization.[11] In passing, we remark that the present formula is also valid for photonic resonators and importantly for the practical case of nanoparticles that are placed on a substrate.

One of the major difficulties to derive such an analytical formula is the establishment of a "mode volume" (or mode normalization) for nanoresonators with strong energy dissipation resulting from either radiation leakage or absorption. The classical mode volume, which is defined with the electromagnetic energy stored in the cavity,[12,13] is accurate only for high-Q dielectric cavities. The theoretical difficulty associated to dielectric cavities with strong leakage has been noted dating back to the early studies of optical microcavities[12-15] and has reemerged in the context of plasmonic nanoresonators,[16,17] in which absorption and dispersion additionally play an important role. Recently, a solid theoretical framework for the computation and normalization of the modes of lossy resonators has been established,[11,18] by transforming the resonance modes, which are morally scattering states, into square-integrable modes by using complex coordinate transforms.

The report is organized as follows. The closed-form expression for the frequency shift and resonance broadening is first derived. Then we explain how the resonance mode can be computed and normalized properly, for instance with the open-source code in ref 19. We further outline the fundamental difference between the present work and previous theoretical works of the same kind.[20-22] A thorough discussion is provided in the Supporting Information. Finally, we test the closed-form-expression accuracy for plasmonic nanoresonators of different sizes and shapes, for perturbations with different shapes, refractive indices and positions with respect to the nanoresonators, and evidence that the formula is highly accurate for a *broad panel* of plasmonic sensors.

**Master equation**. The closed-form expression for $\Delta\tilde{\omega}$ is simple. If we denote by $\tilde{\mathbf{E}}(\mathbf{r})$ the resonance mode (called as quasi-normal mode or QNM, hereafter) of the bare metallic nanoparticle and by $\tilde{\mathbf{E}}_{app}(\mathbf{r}) \approx \tilde{\mathbf{E}}(\mathbf{r})$ a slightly modified version of $\tilde{\mathbf{E}}(\mathbf{r})$ that takes into account local field corrections, the complex-frequency shift $\Delta\tilde{\omega}$ due to a local permittivity change $\Delta\boldsymbol{\varepsilon}(\mathbf{r},\tilde{\omega})$, which may be a tensor for anisotropic media, reads as

$$\Delta\tilde{\omega} = -\tilde{\omega}\iiint_{V_p} \Delta\boldsymbol{\varepsilon}(\mathbf{r},\tilde{\omega})\tilde{\mathbf{E}}_{app}(\mathbf{r}) \cdot \tilde{\mathbf{E}}(\mathbf{r})d^3\mathbf{r}, \qquad (1)$$

where the integral runs over the perturbation volume $V_p$. It will be shown that eq 1 provides accurate predictions for various shapes, sizes, and material properties of the metallic nanoresonator. The main force of eq 1 resides in the fact that, once the QNM is calculated, the shift is known analytically for any shape, size, position or permittivity of the perturbation. Throughout the manuscript, the notation *tilde* is used to refer to QNMs.

At first glance, eq 1 resembles expressions obtained and successfully used in earlier works on resonance shifts for very high-Q dielectric resonators, such as microwave cavities,[23] microspheres,[24] or photonic crystal cavities,[25] and one may question the novelty of the work.

The main difference resides in the integrand used, a $\tilde{\mathbf{E}} \cdot \tilde{\mathbf{E}}$ product instead of a $\tilde{\mathbf{E}} \cdot \tilde{\mathbf{E}}^*$ product and, of course, in the mode normalization. Replacing $\tilde{\mathbf{E}} \cdot \tilde{\mathbf{E}}^*$ by $\tilde{\mathbf{E}} \cdot \tilde{\mathbf{E}}$ is not just a small modification but has profound implications:

- First, on a mathematical perspective, a resonance mode that leaks cannot be normalized with an $\tilde{\mathbf{E}} \cdot \tilde{\mathbf{E}}^*$ product since $\tilde{\mathbf{E}}(\mathbf{r})$ exponentially diverges as

$|\mathbf{r}| \to \infty$.[11] Consistently, normalizations with $\tilde{\mathbf{E}} \cdot \tilde{\mathbf{E}}^*$ products are performed by considering arbitrarily-finite integration domains, in general with a real frequency.[23-25]

- Second, for high-Q photonic resonators, the normalized resonance mode $\tilde{\mathbf{E}}(\mathbf{r})$ is nearly real, $Im(\tilde{\mathbf{E}})/Re(\tilde{\mathbf{E}}) = O(1/Q)$[26] and normalizations with either $\tilde{\mathbf{E}} \cdot \tilde{\mathbf{E}}$ or $\tilde{\mathbf{E}} \cdot \tilde{\mathbf{E}}^*$ on a finite integration domains are largely equivalent when estimating the frequency shift $Re(\Delta\tilde{\omega})$. However, $\tilde{\mathbf{E}} \cdot \tilde{\mathbf{E}}$ and $\tilde{\mathbf{E}} \cdot \tilde{\mathbf{E}}^*$ products provide distinctive predictions for $Im(\Delta\tilde{\omega})$. For instance for a *dielectric perturbation* ($\Delta\varepsilon$ real), the peak broadening $-2Im(\Delta\tilde{\omega})$ with $\tilde{\mathbf{E}} \cdot \tilde{\mathbf{E}}^*$ is always proportional to $Im(\tilde{\omega})$ with a proportionality factor that is real (as can be immediately seen by replacing $\tilde{\mathbf{E}}_{app}$ by $\tilde{\mathbf{E}}^*$ in eq 1), regardless of the mode profile and the perturbation positions. This makes no sense, and consistently, peak-broadening predictions with analytical formulae are essentially absent in the literature on high-Q dielectric cavities, to our knowledge. This difficulty is removed by the formula using $\tilde{\mathbf{E}} \cdot \tilde{\mathbf{E}}$ products. As will be shown in the following, peak broadenings are accurately predicted by eq 1 for low-Q plasmonic resonators. Further work should be undertaken to evaluate the accuracy of eq 1 for predicting the peak broadening of high-Q photonic resonators, but this evaluation remains out of the scope of the present contribution that focuses on plasmonic nanoresonators.

- Third, the phase of the resonant mode, which is ignored by $\tilde{\mathbf{E}} \cdot \tilde{\mathbf{E}}^*$ products matters in eq 1. For tiny perturbations, the phase variation over the spatial extent of the perturbation (i.e., phase-retardation effect) can be neglected, *but for larger perturbations covering the entire surface of the resonator (see Fig. 1c), phase-retardation effect cannot be neglected and should be incorporated in the integral over the perturbation volume*. Phase-retardation effects are strikingly important for predicting light backscattering in optical waveguides in the presence of small fabrication imperfections for instance.[27]

- Fourth, metallic resonators present particular challenges. Because the Q-factor is typically in the range between 10 and 100, $Im(\tilde{\mathbf{E}})$ cannot be neglected, and normalizations with either $\tilde{\mathbf{E}} \cdot \tilde{\mathbf{E}}$ or $\tilde{\mathbf{E}} \cdot \tilde{\mathbf{E}}^*$ provide distinct predictions for both $Re(\Delta\tilde{\omega})$ and $Im(\Delta\tilde{\omega})$. Additionally, since the field around the nanoparticle in free space diverges much faster than for dielectric cavities, the normalization issue becomes critical.[14,17] Beyond the quasi-static limit, the field outside the nanoparticle matters critically for normalization and we will see that eq 1 provides accurate predictions for both $Re(\Delta\tilde{\omega})$ and $Im(\Delta\tilde{\omega})$.

- Finally, to further evidence the predictive force of the present formula, in the supporting information, we make a thorough comparison of the present approach with a recent representative theoretical work of the same kind,[22] in which the relevant "mode" is calculated at the real-valued resonance frequency and the normalization involves $\tilde{\mathbf{E}} \cdot \tilde{\mathbf{E}}^*$ in the quasi-static approximation. Consistently with ref 22, we find that the quasi-static formula accurately predicts $Re(\Delta\tilde{\omega})$ for very small ($<\lambda/10$) resonators. However we find that the quasi-static formula fails in predicting $Im(\Delta\tilde{\omega})$ for any resonator size and rapidly becomes inaccurate for predicting $Re(\Delta\tilde{\omega})$ as the resonator size is increased towards realistic values that are large enough to give sufficient scattering in plasmonic sensing technologies.[6,9] In contrast, these severe limitations are not encountered with our eq 1, evidencing the importance of our universal and robust $\tilde{\mathbf{E}} \cdot \tilde{\mathbf{E}}$ treatment (including computation and normalization) of low-Q resonators.

**Derivation of the master equation.** To derive eq 1, let us start by considering two eigensolutions of source-free Maxwell's equations. The first solution corresponds to the resonance mode of the bare metallic nanoparticle, $\nabla \times \tilde{\mathbf{E}} = -i\tilde{\omega}\boldsymbol{\mu}(\tilde{\omega})\tilde{\mathbf{H}}$ and $\nabla \times \tilde{\mathbf{H}} = i\tilde{\omega}\boldsymbol{\varepsilon}(\tilde{\omega})\tilde{\mathbf{E}}$, denoted as $(\tilde{\mathbf{E}}, \tilde{\mathbf{H}})$ with an eigenfrequency $\tilde{\omega}$, see Fig. 1a. The second solution with an eigenfrequency $\tilde{\omega}'$, $\nabla \times \tilde{\mathbf{E}}' = -i\tilde{\omega}'\boldsymbol{\mu}(\tilde{\omega}')\tilde{\mathbf{H}}'$ and $\nabla \times \tilde{\mathbf{H}}' = i\tilde{\omega}'(\boldsymbol{\varepsilon}(\mathbf{r},\tilde{\omega}') + \Delta\boldsymbol{\varepsilon}(\mathbf{r},\tilde{\omega}'))\tilde{\mathbf{E}}'$, represents the resonance mode $(\tilde{\mathbf{E}}', \tilde{\mathbf{H}}')$ of the nanoresonator dressed by the perturbation (i.e., the permittivity change) $\Delta\boldsymbol{\varepsilon}(\mathbf{r},\tilde{\omega}')$, see Figs. 1b and 1c. Applying the Green-Ostrogradski formula to the vector $\tilde{\mathbf{E}}' \times \tilde{\mathbf{H}} - \tilde{\mathbf{E}} \times \tilde{\mathbf{H}}'$, we obtain

$$\iint_{\Sigma} \left( \tilde{\mathbf{E}}' \times \tilde{\mathbf{H}} - \tilde{\mathbf{E}} \times \tilde{\mathbf{H}}' \right) \bullet d\mathbf{S} =$$
$$-i\iiint_{\Omega} \left\{ \tilde{\mathbf{E}} \bullet [\tilde{\omega}\boldsymbol{\varepsilon}(\tilde{\omega}) - \tilde{\omega}'(\boldsymbol{\varepsilon}(\tilde{\omega}') + \Delta\boldsymbol{\varepsilon}(\tilde{\omega}'))]\tilde{\mathbf{E}}' - \tilde{\mathbf{H}} \bullet [\tilde{\omega}\boldsymbol{\mu}(\tilde{\omega}) - \tilde{\omega}'\boldsymbol{\mu}'(\tilde{\omega}')]\tilde{\mathbf{H}}' \right\} d^3\mathbf{r}$$
(2)

where $\Sigma$ is an arbitrary closed surface defining a volume $\Omega$. In earlier work,[11] it was shown that the volume integral in eq 2 can be evaluated over the entire space, provided that the entire space is decomposed into two sub-domains $\Omega_1$ and $\Omega_2$, with $\Omega_1$ being a finite-volume real space that contains the metallic nanoparticle and $\Omega_2$ a surrounding space that can be implemented with perfectly-matched layers (PMLs), which transform the exponentially-diverging QNMs in real space into square-integrable modes, with an exponential decay in $\Omega_2$, see details in ref 11. Because of the exponential decay, the surface integral on the left-side of eq 2 becomes null, and assuming that $\Delta\tilde{\omega}$ is small so that we may use a first order expansion of the permittivity and permeability for $\omega \approx \tilde{\omega}$, eq 2 becomes

$$\frac{\Delta\tilde{\omega}}{\tilde{\omega}} = \frac{\tilde{\omega}' - \tilde{\omega}}{\tilde{\omega}} = -\frac{\iiint_{V_p} \Delta\boldsymbol{\varepsilon}(\mathbf{r},\tilde{\omega})\tilde{\mathbf{E}}'(\mathbf{r})\cdot\tilde{\mathbf{E}}(\mathbf{r})d^3\mathbf{r}}{\iiint_{\Omega} \left\{ \tilde{\mathbf{E}}(\mathbf{r}) \cdot \frac{\partial[\omega\boldsymbol{\varepsilon}(\mathbf{r},\omega)]}{\partial\omega} \tilde{\mathbf{E}}'(\mathbf{r}) - \tilde{\mathbf{H}}(\mathbf{r}) \cdot \frac{\partial[\omega\boldsymbol{\mu}(\mathbf{r},\omega)]}{\partial\omega} \tilde{\mathbf{H}}'(\mathbf{r}) \right\} d^3\mathbf{r}}, \quad (3)$$

where $\Delta\boldsymbol{\varepsilon}(\mathbf{r},\omega) = \boldsymbol{\varepsilon}_p(\mathbf{r},\omega) - \boldsymbol{\varepsilon}_b(\mathbf{r},\omega)$ with $\boldsymbol{\varepsilon}_p$ and $\boldsymbol{\varepsilon}_b$ the permittivities of perturbation and background medium. It is noteworthy that eq 3 is *exact* up to a first order in $\Delta\tilde{\omega}$.

To calculate the frequency shift $\Delta\tilde{\omega}$ using the sole knowledge of the unperturbed mode $\tilde{\mathbf{E}}$, we need to eliminate the unknown perturbed QNM $\tilde{\mathbf{E}}', \tilde{\mathbf{H}}'$. In the denominator, a very accurate assumption consists in considering that the perturbation modifies the QNM field distribution only locally in a volume approximately equal to $V_p$, and that the error induced by replacing $\tilde{\mathbf{E}}', \tilde{\mathbf{H}}'$ by $\tilde{\mathbf{E}}, \tilde{\mathbf{H}}$ into the denominator of eq 3 is negligible. Then using the QNM normalization $\iiint_{\Omega} \left\{ \tilde{\mathbf{E}} \cdot \frac{\partial[\omega\boldsymbol{\varepsilon}]}{\partial\omega} \tilde{\mathbf{E}} - \tilde{\mathbf{H}} \cdot \frac{\partial[\omega\boldsymbol{\mu}]}{\partial\omega} \tilde{\mathbf{H}} \right\} d^3\mathbf{r} = 1$,[11] we obtain

$$\Delta\tilde{\omega} = -\tilde{\omega}\iiint_{V_p} \Delta\boldsymbol{\varepsilon}(\mathbf{r},\tilde{\omega})\tilde{\mathbf{E}}'(\mathbf{r})\cdot\tilde{\mathbf{E}}(\mathbf{r})d^3\mathbf{r}. \quad (4)$$

To go further, we should estimate $\tilde{\mathbf{E}}'$ from $\tilde{\mathbf{E}}$. A crude approximation would consist in simply making $\tilde{\mathbf{E}}' \equiv \tilde{\mathbf{E}}$ in eq 4. However, we rather consider a better approximation, based on local-field corrections, $\tilde{\mathbf{E}}' \equiv \tilde{\mathbf{E}}_{app}$ as explained in the next Section and finally obtain the master equation eq 1.

We remark that the QNM normalization relies on an analytical continuation in the complex space and can be implemented with PMLs in the domain $\Omega_2$. However a much simpler and completely general method has been reported in ref 18. The method, which can be easily implemented with virtually all numerical Maxwell solvers, calculates normalized QNMs in a few (4-5) iterations that require a few minutes CPU-times. It is this method that we adopt hereafter for the numerical examples, using an open-source COMSOL code.[19]

In the supporting information, we provide a thorough evaluation of errors that occur when replacing $\tilde{\mathbf{E}}'$ by $\tilde{\mathbf{E}}_{app}$ or $\tilde{\mathbf{E}}$ in eq 3. We find that the error made in the denominator is completely negligible and that the dominant error that limits the accuracy of the master equation results from the replacement of $\tilde{\mathbf{E}}'$ in the numerator. In this regard, the error evaluation shows that much better accuracy is achieved with local-field corrections.

**Local field corrections**. In the following, we will consider two types of perturbations: a tiny nano-object placed in the near field of the metallic nanoresonator, see Fig. 1b, and a thin shell surrounding a metallic nanoresonator, see Fig. 1c. The former is important for nanoparticle detection for instance [8,10,28-32] and the latter for biosensing applications[32-34] or gas sensing[35]. For the former case, we adopt the polarizability tensor $\alpha$ of the nano-object to approximate $\tilde{\mathbf{E}}'$ from $\tilde{\mathbf{E}}$, and assume that the perturbed QNM-field $\tilde{\mathbf{E}}'(\mathbf{r})$ inside the nano-object (for $\mathbf{r} \in V_p$) is proportional to the unperturbed QNM-field $\tilde{\mathbf{E}}(\mathbf{r})$[36]

$$\tilde{\mathbf{E}}_{app}(\mathbf{r}) = \boldsymbol{\alpha} \varepsilon_b \tilde{\mathbf{E}}(\mathbf{r}) / \left[ V_p \Delta \boldsymbol{\varepsilon}(\mathbf{r}, \tilde{\omega}) \right]. \tag{5a}$$

The polarizability tensor can be calculated for any particle. For small spheres that will be tested numerically, it takes a simple form $\alpha = 4\pi R^3 (\varepsilon_p - \varepsilon_b)/(\varepsilon_p + 2\varepsilon_b)$. For the shell case, we again assume that the perturbed QNM-field $\tilde{\mathbf{E}}'(\mathbf{r})$ inside the shell (for $\mathbf{r} \in V_p$) is proportional to the unperturbed QNM-field $\tilde{\mathbf{E}}(\mathbf{r})$, with a proportionality factor settled by the field boundary conditions at the nanoresonator-shell interface

$$\tilde{\mathbf{E}}^{(T)}_{app}(\mathbf{r}) \approx \tilde{\mathbf{E}}^{(T)}(\mathbf{r}) \text{ and } \tilde{\mathbf{E}}^{(N)}_{app}(\mathbf{r}) \approx \varepsilon_b/\varepsilon_p \tilde{\mathbf{E}}^{(N)}(\mathbf{r}),  \quad (5b)$$

where the superscripts (T) and (N) refer to tangential and normal field components. This modified first-order Born approximation is known to be very accurate for small shell thicknesses, see for instance a related theoretical study on slow light propagation in photonic-crystal waveguides [27] in the presence of small fabrication imperfections.

**Quantitative numerical tests of the master equation.** For the numerical tests, we consider three types of metallic nanoresonators: a cylindrical gold nanorod (radius $R$ = 15 nm and length $L$ = 90 nm), a dimer composed by two identical cylindrical gold nanorods ($R$ = 15 nm and $L$ = 90 nm, gap size 15 nm) and a gold nanocone ($L$ = 100 nm and $R$ = 27.5 nm at bottom). The apexes of the nanorods and the nanocone are rounded (the nanorod apexes are hemispheres with the same radius as the cylinder and the nanocone apex is rounded with a radius of curvature $R$ = 7.5 nm). The nanoresonators are assumed to be surrounded by water ($\varepsilon_b$ = 1.77). The z-component of the electric field of the fundamental QNMs are shown in Fig. 2. Hot spots in water show up with a blue color.

To test the accuracy of eq 1, we compare the resonance shift $\Delta\tilde{\omega}$ predicted with eq 1 to the exact shift value $\Delta\tilde{\omega}_{exact}$ computed as the difference of the eigenwavelengths of the perturbed and the unperturbed QNMs, both obtained with the open-source COMSOL code[19] with two independent computations. The fully-vectorial approach adopted here to obtain $\Delta\tilde{\omega}_{exact}$ is equivalent to the usual approach ( more details can be found in the discussion about the last example), which consists in comparing the spectral responses (e.g., spectra of the absorption or extinction) of a perturbed resonator and an unperturbed resonator, but it is more efficient from numerical perspectives and is also more accurate for estimating small changes of the resonance width.

Figure 3a shows the wavelength shifts induced by the presence of a protein nanosphere ($\varepsilon_p$ = 2.25) located in the hot spots of the three metallic nanoresonators at a 0.5 nm separation distance from the metal surface. These situations are encountered in single-molecule sensing experiments, where the analyte can be chemically adsorbed to the nanoresonator surface with an analyte-metal spacing of $\approx$ 1 nm.[9,10] The best sensitivity is obtained for the nanocone (brown-thick curve). This is straightforwardly deduced from the hot-spot intensities in Fig. 2, since the brighter spot is achieved at the nanocone apex. This underlines the importance of a proper disposal of QNM normalization. The dimer performance is tested for two nanosphere locations in the gap, either 0.5 nm below the upper arm (black curve) or exactly at the gap center (red curve). On the whole, the dimer performance is better than that of its constituting element, the single nanorod (blue curve); remarkably, it becomes as sensitive as the nanocone. This is again easily understood from Fig. 2: close to the nanocone apex, the QNM near-field is the strongest one, but it also the most confined so that a balance is achieved for relatively large nanospheres. In general, the agreement between the analytical predictions obtained with eq 1 and the fully-vectorial calculations is excellent in Fig. 3a. We have performed additional tests for nanorods with different dimensions (not shown) and similar agreement was achieved; we found that as the size of the nanorod apex increases, the sensitivity decreases.

Figure 3b reports additional tests performed for perturbing nanospheres of high permittivity $|\varepsilon_p|$, such as those used for the sensing applications in which target analytes are labeled by high-permittivity nanoparticles.[28,30] Two spherical nanoparticles with radius $R$ = 3nm are considered; one is made of gold ($\varepsilon_p(\tilde{\omega})$ = −20.45+0.81$i$ at resonance) and the other one of silicon ($\varepsilon_p$ = 12.25). The shifts are calculated as a function of the separation distance to the nanorod apex, see inset. Again the agreement is excellent, at least for separation distances $S$ larger than 1 nm. For $S$ < 1 nm, the local field correction of eq 5a is no longer valid. As shown by an inspection of the near-field distribution $\tilde{\mathbf{E}}'(\mathbf{r})$ of the perturbed nanorod, the failure of the local field correction can be ascribed to the emergence of strong gap resonance [37,38] confined between the nanorod apex and the high-permittivity nanosphere for $S$ < 1 nm. We remark that though the change of the resonance width, in terms of $2Im(\Delta\tilde{\lambda})$, is small, the analytical formula is accurate enough to capture this subtle information.

Figure 4 summarizes the tests made for protein ($\varepsilon_p$ = 2.25) shells adsorbed on a gold nanorod ($R$ = 10 nm, $L$ = 80 nm and $\tilde{\lambda}$ = 792.55−28.02$i$ nm) in aqueous environment ($\varepsilon_b$ = 1.769). This corresponds to sensing techniques, for which a self-assembled-monolayer analyte envelopes a metallic nanoparticle.[6,32,33] In Fig. 4a, predictions obtained with eq 1 containing the local field correction of eq 5b, are plotted as solid curves; again quantitative agreement with fully-vectorial calculations (circles) is achieved even for shell thicknesses as large as a few nanometers. The calculation indicates a nearly linear dependence of $Re(\Delta\tilde{\lambda})$ and $Im(\Delta\tilde{\lambda})$ on the shell thickness, consistently with experimental observations.[16] In Fig. 4b, we show the scattering cross-section spectra of the bare nanorod (blue) and the perturbed nanorod (5-nm shell, red). The spectra are calculated using COMSOL multiphysics. By fitting with a Lorentzian function, we find a red-shift of 36 nm and a resonance broadening of ≈3.6 nm. These values, which are inferred from spectrum characteristics that are typically encountered in real experiments, are fully consistent with the predictions of the closed-form expression $Re(\Delta\tilde{\lambda})$ ≈ 35.1 nm and $-2Im(\Delta\tilde{\lambda})$ ≈ 3.4 nm, offering a posteriori validation of the pure

QNM approach (namely, the fully-vectorial computations of frequency shift) used for testing the accuracy of the prediction of closed-form expression in Figs. 3a, 3b and 4a.

Finally, throughout this paper we consider single mode cases, where the spectral responses of the metallic nanoresonators are driven solely by a single resonance mode and the scattering or absorption spectra are basically *Lorentzian* just as the ones in Fig. 4b. In multimode cases, on the other hand, intricate Fano-like spectral lineshapes [11,39-42] can be observed because the perturbation may affect several QNMs, leading to complicated deformation in the spectral lineshapes. One may extend the current perturbation approach to these cases,[11] by considering the interaction of the perturbation with each QNM and obtain more physical insight. However, in this case one must be cautious with the concepts of *frequency shift* and *peak broadening*. Additionally, we consider nanospheres, as isolated perturbing nanoparticles, with a well-known polarizability to test the analytical formula; however, in general one should ideally adopt proper polarizabilities for perturbing nanoparticles with different shapes to determine the local field correction.[36]

**Conclusion**. We have derived a closed-form expression to predict the localized-plasmon-resonance shift and broadening of metallic nanoresonators induced by tiny perturbations of the resonator near-fields. The successful derivation benefits from a recent theoretical advance in the computation and normalization of the resonance modes of lossy nanoresonators. Verification of the accuracy of closed-form expression for various sensing configurations has been performed by comparison with fully-vectorial calculations of the complex eigenfrequencies of the perturbed nanoresonators. We emphasize that the present approach is not stringently restricted by the size or shape of the nanoresonators or perturbations and it may be used for nanoresonators laying on a substrate. Finally it is worth emphasizing that, provided that one is equipped with a Maxwell QNM solver such as the one we have used in this work, the resonance shift is known analytically. It would be interesting to consider extension of the present work to nanoresonators that operate by combining several resonances, such as complex systems sustaining Fano resonances.[43]

ASSOCIATED CONTENT
**Supporting Information**. (1) Comparison of the present master equation with a quasi-static formula recently published.[22] (2) Analysis of the errors resulting from the approximations made to derive the master equation. This material is available free of charge via the Internet at http://pubs.acs.org.

AUTHOR INFORMATION
**Corresponding Author**
*E-mail: philippe.lalanne@institutoptique.fr

**Notes**
The authors declare no competing financial interest.

**Acknowledgment.** J.Y. and P.L. thank M. Perrin, C. Sauvan and J. P. Hugonin for fruitful discussions and help. Part of the study was carried out with financial support from "the Investments for the future" Programme IdEx Bordeaux–LAPHIA (ANR-10-IDEX-03-02). H.G. thanks the ERC (Complexplas), the BMBF, DFG, and BW Stiftung for support.

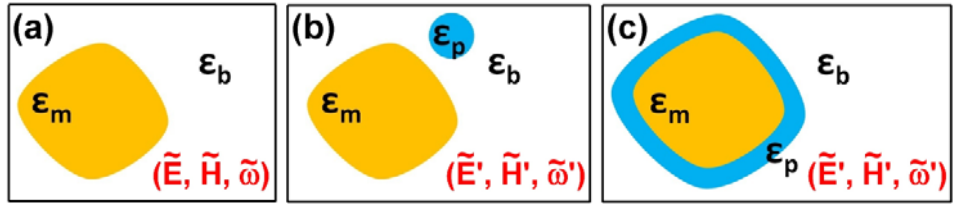

**Figure 1. Schematic showing unperturbed and perturbed resonant metallic nanoparticles**. (**a**) Bare metallic nanoparticle supporting a quasi-normal mode (QNM) $(\tilde{\mathbf{E}}, \tilde{\mathbf{H}})$ at eigenfrequency $\tilde{\omega}$. (**b**) and (**c**) Metallic nanoparticle perturbed by a small nanosphere or a thin shell, supporting a perturbed QNM $(\tilde{\mathbf{E}}', \tilde{\mathbf{H}}')$ at eigenfrequency $\tilde{\omega}'$. The permittivities of the metal, background medium and perturbation are denoted by $\varepsilon_m$, $\varepsilon_b$ and $\varepsilon_p$, respectively.

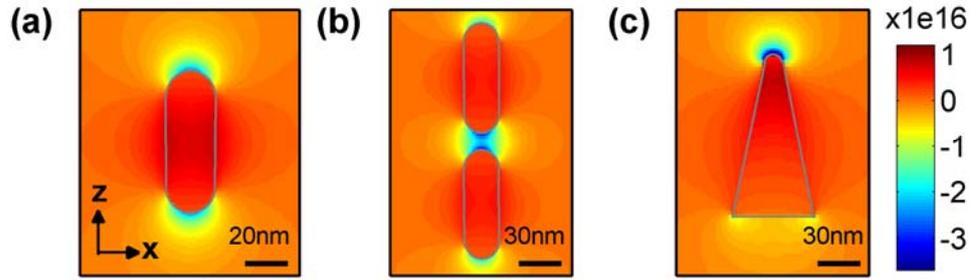

**Figure 2. The electric field distribution Re($\tilde{E}_z$) of the normalized fundamental QNMs supported by three gold nanoresonators.** (**a**) a single nanorod, (**b**) a nanorod dimer and (**c**) a nanocone. The nanoresonators are placed in water ($\varepsilon_b$ = 1.77) and their dimensions is given in the main text. The associated eigenwavelengths, computed with the open-source COMSOL software in ref 19, are $\tilde{\lambda}$ = $2\pi c/\tilde{\omega}$ = 691.52–30.94$i$ nm, 756.44 – 47.12$i$ nm and 805.31 – 40.28$i$ nm, respectively. In the simulation, a Drude model $\varepsilon_m = 1 - \omega_p^2/(\omega^2 - i\omega\Gamma)$ is adopted for the relative permittivity of gold with $\omega_p$ = 1.26x10$^{16}$ s$^{-1}$ and $\Gamma$ = 1.41x10$^{14}$ s$^{-1}$.

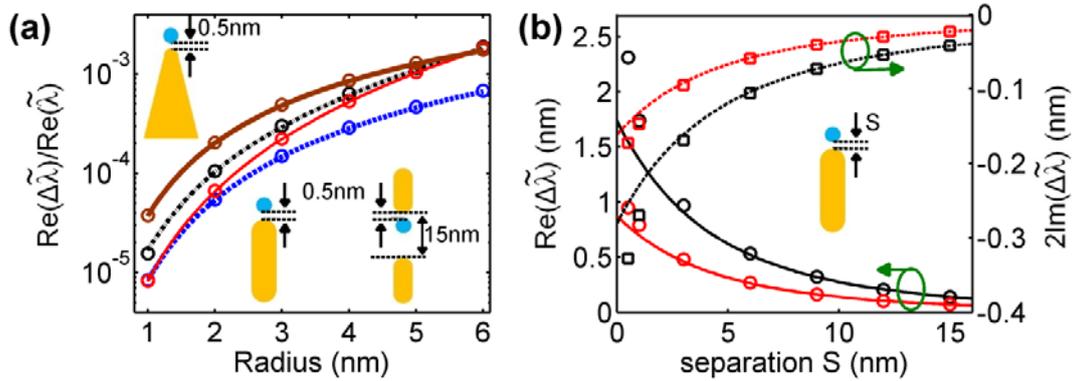

**Figure 3. Resonance shifts of plasmonic nanostructures due to a perturbing nanosphere in aqueous environment ($\varepsilon_b$ = 1.77).** (a) $Re(\Delta\tilde{\lambda})/Re(\tilde{\lambda})$ as a function of the nanosphere (protein, $\varepsilon_p$ = 2.25) radius for various nanoresonators. The sphere is placed 0.5 nm above the apex of the nanorod (blue dashed) or nanocone (brown thick). For the dimer, the nanosphere resides in the gap, at 0.5 nm below the upper arm (black dotted-dash) or at the gap center (red thin). (b) $Re(\Delta\tilde{\lambda})$ and $2\cdot Im(\Delta\tilde{\lambda})$ as a function of the separation distance S between the nanorod and the perturbing nanosphere of radius R = 3 nm, for nanospheres made of gold (black curves) and silicon (red curves, $\varepsilon_p$ = 12.25). In (a) and (b), circle or square marks are obtained with fully-vectorial calculation and curves are predicted with eq 1.

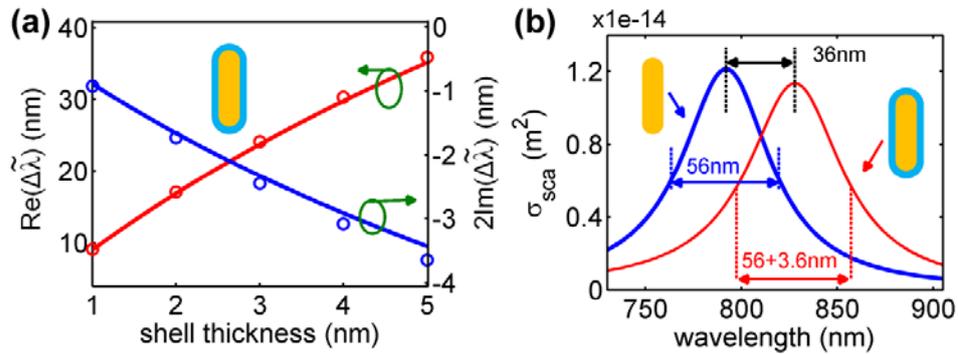

**Figure 4. Resonance shift induced by the formation of a thin dielectric shell around a plasmonic nanorod in aqueous environment ($\varepsilon_b$ = 1.77).** (a) $Re(\Delta\tilde{\lambda})$ (red) and $2\cdot Im(\Delta\tilde{\lambda})$ (blue) as a function of the shell thickness. Fully-vectorial calculations and analytical predictions are represented by circles and solid curves. (**b**) Calculated scattering cross-section $\sigma_{sca}$ as a function of the wavelength, for a bare nanorod (blue) and the same nanorod covered by a 5-nm shell (red). In (a) and (b), the shell permittivity is $\varepsilon_p$ = 2.25 and the rod radius and length are $R$ = 10 nm and $L$ = 80 nm.

**Supporting Information**

# Simple analytical expression for the peak-frequency shifts of plasmonic resonances for sensing

Jianji Yang[1], Harald Giessen[2] and Philippe Lalanne*[1]

[1] Laboratoire Photonique Numérique et Nanosciences, Institut d'Optique d'Aquitaine, Univ. Bordeaux, CNRS, 33405 Talence, France.
[2] 4th Physics Institute and Research Center SCoPE, University of Stuttgart, Pfaffenwaldring 57, 70550 Stuttgart, Germany.

* E-mail: Philippe.lalanne@institutoptique.fr

**Content of the document**
(1) Comparison of the present master equation eq 1 in the main text with a quasi-static formula recently published [1]
(2) Analysis of the errors resulting from the approximations made to derive the master equation eq 1 (main text)

# 1. Comparison of the present Master Equation with the quasi-static Formula

A formula for predicting frequency shifts of resonances of plasmonic nanoresonators was recently reported in eq 11 in Ref [1]. With the same notation as in Ref [1], it reads as

$$\Delta\omega_m = -\alpha_{NP}\frac{d\omega}{d\varepsilon_{ca}}\frac{|\mathbf{E}_m(\mathbf{r}_{NP})|^2}{\iiint_{\text{cavity}}|\mathbf{E}_m(\mathbf{r})|^2 d^3\mathbf{r}}, \qquad (S1)$$

where $\alpha_{NP}$ and $\varepsilon_{ca}$ denote the polarizability of the perturbation (assumed to be infinitely small and placed at $\mathbf{r}_{NP}$) and the permittivity of the metallic resonator respectively. The integral in the denominator runs over the metallic resonators. Referring to ref [1], the field $\mathbf{E}_m$ is defined as the field scattered by the metallic resonator under plane-wave illumination at the *real-valued* resonance frequency $\omega_m$. Strictly speaking this field is not the same as the resonance mode (the quasi-normal mode) that is used in our work, which is defined with a complex frequency. Hereafter, following [1] we will use the scattered field $\mathbf{E}_m$ to evaluate eq (S1) and refer to eq (S1) as *quasi-static formula*, since it is derived using quasi-static approximation.

A weakness of the quasi-static theory is that it is valid only at deep sub–λ scales. In this Section, using fully-vectorial calculations, we compare the accuracy of our present master equation (eq 1 main text) and the quasi-static formula, for metallic resonators with progressively increasing sizes.

However, before starting any comparison, we first check that we are correctly implementing eq (S1) by reproducing some of the results in [1]. For that purpose, we consider three resonators, a silver ellipsoid (40-nm long axis, 14-nm short axes), a silver nanorod dimer (arm size 28 nm x 10 nm x 10nm, 10-nm gap), and a silver split-ring resonator (10 nm x 10 nm wire cross-section, 32 nm x 30 nm outer dimension of the ring, 10 nm gap). These geometries are the same as those used in Fig. 2 in [1]. For each geometry, we calculate the exact complex frequency shift $\Delta\tilde{\omega}_{\text{exact}}$ as the difference between the complex eigenfrequencies of the perturbed and the unperturbed one (see the main text) and define $\Delta\omega_{\text{exact}} = Re(\Delta\tilde{\omega}_{\text{exact}})$. Following [1], we consider a tiny silicon nanosphere as the perturbation. We calculate the scattered field $\mathbf{E}_m$ is with COMSOL and denote the predicted shift by $\Delta\omega_{\text{predict}}$.

In Fig. S1, we plot $\Delta\omega_{\text{exact}}$ as a function of $\Delta\omega_{\text{predict}}$ for the three resonators and for different perturbation locations (see the caption). As in Ref [1], we plot the line $\Delta\omega_{\text{exact}} = 1.07\Delta\omega_{\text{predict}}$ (black line). The good agreement with Fig. 2b of Ref [1] makes us confident with our implementation of the quasi-static formula.

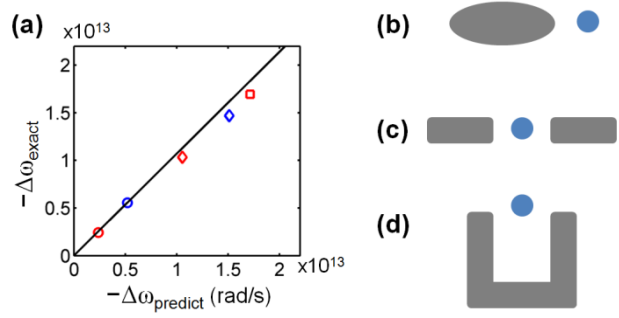

**Figure S1. Test of our capability to correctly evaluate eq (S1) by reproducing Fig. 2b in [1].** Three silver resonators in air taken from Ref [1] are used. The perturbation (blue dot) is a silicon nanosphere (2.5-nm radius, $n$ = 3.5) like in Ref [1]. (**a**) Peak frequency shifts for ellipsoid (circles), nanorod dimer (square) and SRR (diamonds). The black line has a slope of 1.07. (**b**) Sketch of the ellipsoid. The nanosphere is along the long-axis, with 3-nm (red circle) and 5-nm (blue circle) separation respectively. (**c**) Sketch of the nanorod dimer. The nanosphere is in the gap center. (**d**) Sketch of the SRR. The nanosphere is placed along the central axis, being 15 nm (red diamond) and 20 nm (blue diamond) above the bottom arm. A Drude model $\varepsilon_m = 4 - \omega_p^2/(\omega^2 - i\omega\Gamma)$ is adopted for the relative permittivity of silver with $\omega_p$ = 1.4x10$^{16}$ s$^{-1}$ and $\Gamma$ = 3.2x10$^{13}$ s$^{-1}$.

**Comparison.** Reinforced by the agreement, we systematically compare the accuracy of the present master equation, based on quasi-normal mode (QNM) theory, and that of the quasi-static formula. For the comparison, we consider metallic resonators perturbed by a nanosphere. The resonators are silver ellipsoids (taken from [1]), gold nanorods and nanorod dimers (both taken from the main text). The resonator sizes are gradually increased, using a *scaling factor* (SF), from *deep* sub–λ *scale* (~λ/15) (similar to those in [1] and in Fig. S1) to ~λ/4. This size range is typically encountered in plasmonic sensing applications.

In fact, the quasi-static formula predicts complex-valued resonance shifts $\Delta\tilde{\lambda}$ as the metal permittivity $\varepsilon_{ca}$ is complex and the perturbation permittivity (taken into account via $\alpha_{NP}$) may be complex as well. So we provide the comparison for both $Re(\Delta\tilde{\lambda})$ and $Im(\Delta\tilde{\lambda})$, although no predictions of $Im(\Delta\tilde{\lambda})$ are reported in Ref [1]. We emphasize that, though in plasmonic sensing the peak broadening is rarely considered, both peak broadening and shift are fundamental effects associated to a perturbed resonance, thus they should be treated together into a theoretical work, just like in previous works on high-Q RF and photonic cavities [2-4].

The resonance shifts, obtained with fully-vectorial calculations ($\Delta\tilde{\lambda}_{exact}$), with the quasi-static formula ($\Delta\tilde{\lambda}_{static}$) and with the master equation ($\Delta\tilde{\lambda}_{QNM}$), are compared in Figs. S2 (ellipsoid), S3 (nanorod), and S4 (dimer). We also compare the relative errors, defined as the normalized difference between the exact and approximated values. $\Delta\tilde{\lambda}_{\text{exact}}$ is obtained as the difference between the complex-valued eigenwavelengths of the perturbed and the unperturbed QNMs. To provide a comparison as accurate as possible, consistently with eq (S1) and the work in [1], when using the master equation, we assume that the field in the perturbation is uniform and given by the field at the perturbation center.

Figs. S2, S3, and S4 evidence that the quasi-static formula cannot predict $Im(\Delta\tilde{\lambda})$ at all, and it is accurate for $Re(\Delta\tilde{\lambda})$ only at deep sub–$\lambda$ scale, typically for SF ≤ 1.5. Such a small value sets a severe upper bound on the characteristic transverse size that can be considered with accuracy (< $\lambda$/10); for SF = 1.5, the surfaces of the nanorod and the ellipsoid are 2500 nm$^2$ and 3200 nm$^2$, respectively. They correspond to nanospheres with 25~30 nm diameters, being one or two orders of magnitude smaller than the typical surface areas used in biosensing applications [5,6]. Additionally note that absorption and scattering cross-sections become comparable when the nanospheres are about 60 nm in diameter for silver nanospheres and 80 nm in diameter for gold nanospheres.

On the other hand, predictions of both $Re(\Delta\tilde{\lambda})$ and $Im(\Delta\tilde{\lambda})$ made by the master equation are highly accurate at least up to SF = 4, with a relative error below 5% that is even decreasing as the nanoparticle sizes increase. We have check that increase of the relative error as the nanoparticle dimensions are scaled down is due to the fact that the QNM field varies over the perturbation for small nanoresonators and the polarizability model used for the comparison does not consider such variations. We have checked that by taking into account the variation as it is done in the main text.

**Conclusion.** The quasi-static formula could be only applicable for ultrasmall metallic resonators (< $\lambda$/10). The QNM-theory-based master equation is not limited by size and might be used for a broader variety of nanoparticles and realistic applications. As QNMs are the truly eigensolutions of Maxwell's equations, without approximations about resonator size, the master equation is valid independent of the resonator size.

Furthermore, concerning practical implementation, the QNM-based master equation and the quasi-static formula both require the computation of the resonance mode profile, and demand very similar computational loads. We also provide a robust numerical protocol for calculating and normalizing QNMs [8,9]. For metallic resonators of regular shapes and sizes (like for the examples shown here), it just consumes a few minutes to obtain a QNM with a low speed computational workstation equipped with a finite element software.

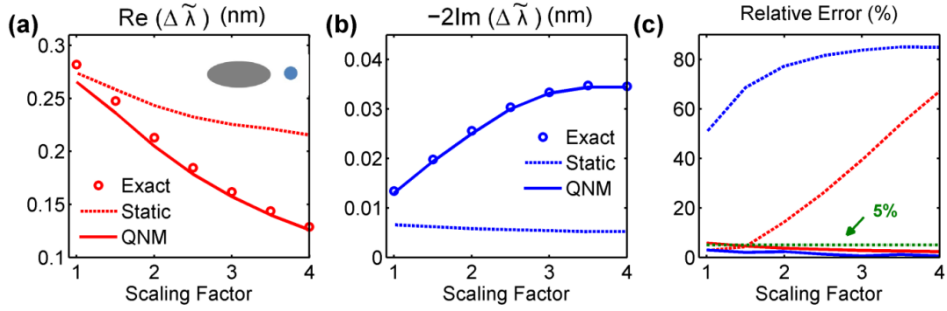

**Figure S2. Resonance shifts of Ag ellipsoids in air perturbed by a silicon nanosphere (2.5-nm radius, *n* = 3.5), with a 5-nm separation (fixed).** (**a**) and (**b**) $Re(\Delta\tilde{\lambda})$ and $-2\cdot Im(\Delta\tilde{\lambda})$, as a function of the scaling factor, obtained with fully-vectorial simulations (circles), the quasi-static formula (dashed, denoted by *static*) and the master equation (solid, denoted by *QNM*). (c) Relative errors, defined as |exact − prediction|/|exact|, for the quasi-static formula (dashed) and the master equation (solid). Red and blue curves correspond to $Re(\Delta\tilde{\lambda})$ and $Im(\Delta\tilde{\lambda})$. The green line marks a 5% relative error. Scaling factor SF = 1 corresponds to the ellipsoid in [1], with 40-nm long axis and 14-nm short axes.

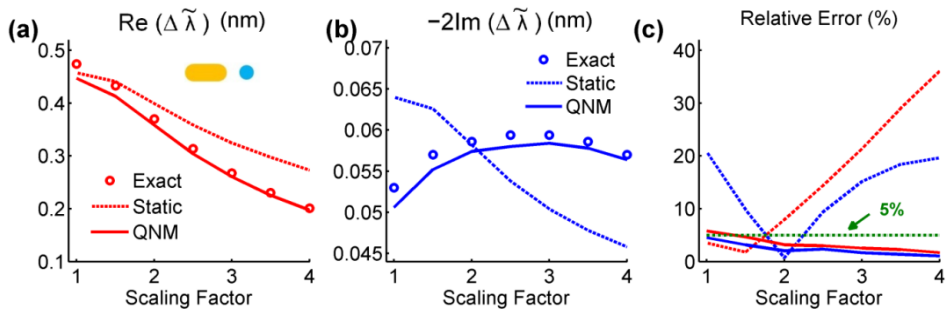

**Figure S3. Same as in Fig. S2 except that gold nanorods are considered.** The nanorods are immersed in water (*n* = 1.33) perturbed by a silicon nanosphere (3-nm radius, *n* = 3.5), with a 6-nm separation (fixed). Scaling factor SF = 1 corresponds to a nanorod with 5-nm radius and 30-nm length, and SF = 3 corresponds to the nanorod considered in Fig. 3 (main text).

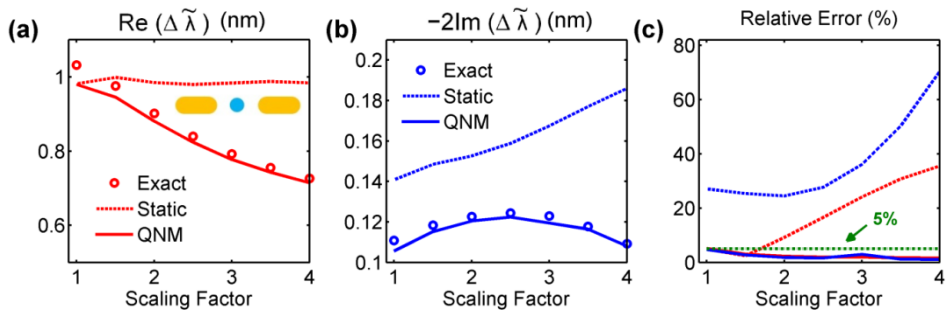

**Figure S4. Same as in Fig. S2 except that gold nanorod dimers are considered.** The dimers are immersed in water (*n* = 1.33) and perturbed by a protein nanosphere (5-nm radius, *n* = 1.5), placed at the gap center. Scaling factor SF = 1 corresponds to a dimer, with 15-nm gap between the two arms (5-nm radius and 30-nm length), and SF = 3 corresponds to the one considered in Fig. 3 (main text). The gap is fixed to be 15 nm for all scaling factors.

**Table S1** lists all the eigenwavelengths $\tilde{\lambda}_m$ of the original QNMs associated to the resonators shown in Figs. S2 – S4.

**Table S1. Eigenwavelengths $\tilde{\lambda}_m$ of Original QNMs** (unit: nm)

| Scaling Factor | Silver Ellipsoid | Gold Nanorod | Gold Nanorod Dimer |
|---|---|---|---|
| 1.0 | 467.09 – 2.61$i$ | 620.75 – 14.77$i$ | 636.82 – 16.18$i$ |
| 1.5 | 476.31 – 4.48$i$ | 632.56 – 16.54$i$ | 660.14 – 19.96$i$ |
| 2.0 | 489.07 – 7.98$i$ | 648.54 – 19.73$i$ | 688.09 – 26.29$i$ |
| 2.5 | 505.25 – 13.43$i$ | 668.32 – 24.52$i$ | 720.31 – 35.36$i$ |
| 3.0 | 524.69 – 20.98$i$ | 691.52 – 30.94$i$ | 756.44 – 47.12$i$ |
| 3.5 | 547.31 – 30.58$i$ | 717.74 – 38.89$i$ | 796.24 – 61.24$i$ |
| 4.0 | 572.97 – 42.02$i$ | 746.68 – 48.26$i$ | 839.51 – 77.53$i$ |

## 2. Analysis of the errors induced by the approximations made to derive the Master Equation

The master equation eq 1 (main text) is derived by applying two approximations into eq 3, which is directly derived by applying Lorentz reciprocity to the original and perturbed modes $(\tilde{\mathbf{E}}, \tilde{\mathbf{H}})$ and $(\tilde{\mathbf{E}}', \tilde{\mathbf{H}}')$ and therefore offers the *exact* shift $\Delta\tilde{\omega}_{exact}$. Hereafter we analyze the impact of the approximations on the prediction accuracy.

For the sake of clarity, we rewrite eq 3 as

$$\frac{\Delta\tilde{\omega}}{\tilde{\omega}} = -\frac{N_{exact}}{D_{exact}}, \qquad (S2)$$

where $N_{exact}$ and $D_{exact}$ respectively denote the numerator and denominator of the right-hand side of eq 3. To derive the master equation that requires the sole knowledge of the original mode $(\tilde{\mathbf{E}}, \tilde{\mathbf{H}})$, we use **two approximations**:

(**i**) in the denominator,

$$D_{exact} = \iiint_{\Omega} \left\{ \tilde{\mathbf{E}}(\mathbf{r}) \cdot \frac{\partial[\omega\varepsilon(\mathbf{r},\omega)]}{\partial\omega} \tilde{\mathbf{E}}'(\mathbf{r}) - \tilde{\mathbf{H}}(\mathbf{r}) \cdot \frac{\partial[\omega\mu(\mathbf{r},\omega)]}{\partial\omega} \tilde{\mathbf{H}}'(\mathbf{r}) \right\} d^3\mathbf{r} \quad , \qquad (S3)$$

we replace $(\tilde{\mathbf{E}}', \tilde{\mathbf{H}}')$ by the normalized original mode $(\tilde{\mathbf{E}}, \tilde{\mathbf{H}})$, thus $D_{exact}$ becomes

$$D_{appr} = \iiint_{\Omega} \left\{ \tilde{\mathbf{E}} \cdot \frac{\partial[\omega\varepsilon]}{\partial\omega} \tilde{\mathbf{E}} - \tilde{\mathbf{H}} \cdot \frac{\partial[\omega\mu]}{\partial\omega} \tilde{\mathbf{H}} \right\} d^3\mathbf{r} = 1;$$

(**ii**) in the numerator,

$$N_{exact} = \iiint_{V_p} \Delta\varepsilon(\mathbf{r},\tilde{\omega})\tilde{\mathbf{E}}'(\mathbf{r}) \cdot \tilde{\mathbf{E}}(\mathbf{r}) d^3\mathbf{r} , \qquad (S4)$$

we replace $\tilde{\mathbf{E}}'$ by the modified version $\tilde{\mathbf{E}}_{app}$ of $\tilde{\mathbf{E}}$.

We evaluate the errors due to approximations (**i**) and (**ii**) for an example shown in Fig. 3b (main text), in which a gold nanorod is perturbed by a gold nanosphere (3-nm radius). The example is selected because the predicted shifts $\Delta\tilde{\omega}_{predict}$ show particularly large (compared to other examples) deviations from the exact shifts $\Delta\tilde{\omega}_{exact}$, especially for small nanorod-nanosphere separation distances (≤ 1 nm).

**Denominator errors.** Figure S5(a) shows the relative error $|D_{appr} - D_{exact}|/|D_{exact}|$ made on the denominator by replacing $D_{exact}$ by $D_{appr}$ as a function of the separation. Actually, the relative error $|D_{appr} - D_{exact}|/|D_{exact}|$ is less than 1% for separations as small as 0.5 nm. Therefore the replacement of $(\tilde{\mathbf{E}}',\tilde{\mathbf{H}}')$ by $(\tilde{\mathbf{E}},\tilde{\mathbf{H}})$ in the denominator, i.e. approximation (**i**), introduces negligible errors, which are comparable to the ratio between the perturbation volume and the mode volume.

**Numerator errors.** For approximation (**ii**), two versions can be applied: (**A**) a crude one $\tilde{\mathbf{E}}_{app} = \tilde{\mathbf{E}}$ and (**B**) $\tilde{\mathbf{E}}_{app} = \alpha\varepsilon_b\tilde{\mathbf{E}}/[V_p \Delta\varepsilon(\mathbf{r},\tilde{\omega})]$ (adopted in the main text), which incorporates local field correction [10]. Let's denote the corresponding numerator as $N_{appA}$ and $N_{appB}$, respectively. Figures S5(b) and S5(c) show $Re(\Delta\tilde{\lambda})$ and $-2Im\cdot(\Delta\tilde{\lambda})$, calculated with combinations $(N_{appA},D_{appr})$ (black squares), $(N_{appB},D_{appr})$ (green triangles) and $(N_{exact},D_{exact})$ (red curve). We have checked that, consistently with Fig. S5(a), predictions obtained with the combination $(N_{exact},D_{appr})$ are superimposed with the exact values and not shown for the sake of clarity. Predictions obtained using $(N_{appA},D_{exact})$ show large errors and evidence the need to apply local field corrections. However $(N_{appB},D_{appr})$ leads to much more accurate predictions. Note that it is this combination that is used to obtain the data in Fig. 3 (main text).

Considering the negligible error in the denominator, the stringent deviation between predictions obtained by $(N_{appB},D_{appr})$ and exact values for small separations (≤1 nm) indicates that local-field corrections cannot offer accurate estimation of $\tilde{\mathbf{E}}'$ for some extreme cases. Careful observation (not shown here) of $\tilde{\mathbf{E}}'$ for separations smaller than 1 nm revealed the formation of an intense gap plasmon, tightly confined between the gold nanorod (nanoparticle) and the gold nanosphere (perturbation).

**Summary.** The evaluation evidences that the major approximation used for deriving the master equation arises from the replacement of $\tilde{\mathbf{E}}'$ by $\tilde{\mathbf{E}}_{app}$ in the numerator of eq 3 (main text) and that it is necessary to use local field correction for high accuracy. In contrast, the error introduced by approximation (**i**), i.e. the replacement of $(\tilde{\mathbf{E}}',\tilde{\mathbf{H}}')$ by $(\tilde{\mathbf{E}},\tilde{\mathbf{H}})$ in the denominator of eq 3, is negligible.

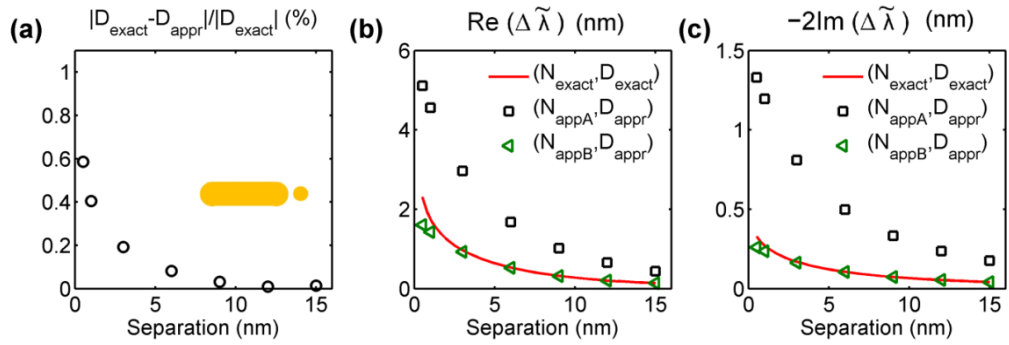

**Figure S5. Errors introduced by applying approximations (i) and (ii) into eq 3 (main text) for the derivation of the master equation (main text).** The errors are estimated for a gold nanorod perturbed by a gold nanosphere (inset) as a function of the nanorod-nanosphere separation. **(a)** $\left|D_{appr} - D_{exact}\right|/\left|D_{exact}\right|$. **(b)** and **(c)** Resonance shifts, $Re(\Delta\tilde{\lambda})$ and $-2\cdot Im(\Delta\tilde{\lambda})$, obtained with combinations $(N_{appA}, D_{appr})$ and $(N_{appB}, D_{appr})$ are shown with black squares and green triangles. The exact value, corresponding to $(N_{exact}, D_{exact})$, is shown with the red curve.